\documentclass[a4paper,12pt]{article}
\usepackage{amssymb}

\usepackage{graphicx}
\usepackage{amsmath}
\usepackage{a4}

\input{tcilatex}

\begin{document}

\title{On the Spin content of the Nucleon}
\author{N. F. Nasrallah\thanks{%
e-mail: nsrallh@cyberia.net.lb} \\
Faculty of Sciense, Lebanese University\\
Tripoli, Lebanon.}
\date{}
\maketitle

\begin{abstract}
A QCD sum rule calculation of Balistky and Ji on the spin content of the
nucleon is done with a different approach to the evaluation of the bilocal
contributions and to the extraction of the nucleon pole residues. The result
obtained is much more numerically stable which puts their conclusion that
more than half of the nucleon spin is carried by gluons on firmer ground.
\end{abstract}

The question of the spin distribution of the nucleon among its constituents
has recently received much attention and it is now realized that a
substantial fraction of the nucleon spin is carried by the quark orbital and
gluon angular momenta \cite{1}

\begin{equation}
\frac{1}{2}=\frac{1}{2}\Delta \Sigma \left( \mu ^{2}\right) +L_{q}\left( \mu
^{2}\right) +J_{q}\left( \mu ^{2}\right)  \label{1}
\end{equation}

Where $\mu ^{2}$ is the scale at which the relevant operators are
renormalized (we shall take here $\mu ^{2}\simeq 1$\textrm{GeV}$^{2}$ and
omit it \ from the notation). The first term in eq.(\ref{1}) has been
measured in polarized deep inelastic scattering \cite{2}. The second and
third terms represent quark orbital and gluon contributions respectively.

Ji \cite{3} has recently shown that the total quark (and hence gluon)
contribution to the nucleon spin is measurable through virtual Compton
scattering in a special kinematic region where single quark scattering
dominates via the study of off forward parton distributions (OFPD) whose
second moment yields the quark contribution.

It was further shown by Vanderhaegen, Guichon and Guidal \cite{4} that hard
electroproduction of photons and mesons on the nucleon also gives access to
the OFPD.

These authors also suggest that preliminary measurements can soon start at
HERMES\ and CEBAF\ which can shed light on the spin content of the proton.

An explicitly gauge invariant decomposition of the angular momentum operator
in QCD was first given by Ji \cite{3}

\begin{equation}
\vec{J}_{\mathrm{QCD}}=\int \left[ \frac{1}{2}\bar{\psi}~\vec{\gamma}~\gamma
_{5}~\psi +\psi ^{+}\vec{x}\times i\vec{D}\,\psi +\vec{x}\times \vec{E}%
\times \vec{B}\right] ~d^{3}x  \label{2}
\end{equation}
This has subsequently been used by Balitsky and Ji \cite{5} to study the
three point function

\begin{equation}
W_{g}^{\mu \nu \alpha }(p)\underset{\lim q\rightarrow 0}{=}\int \exp
(ipx)\exp (iqz)~\langle 0\mid T~\eta (x)~\bar{\eta}(0)~M_{g}^{\mu \nu \alpha
}(z)\mid 0\rangle ~dx~dz  \label{3}
\end{equation}
where $\eta $ is the nucleon field \cite{6}

\begin{equation*}
\eta (x)=\epsilon ^{ijk}(u^{iT}c~\gamma ^{\alpha }~u^{j})~\gamma _{5}~\gamma
_{\alpha }~d^{k}\ \ \ \ 
\end{equation*}
\ \ \ \ \ \ \ \ \ $M^{\mu \gamma \alpha }=T^{\mu \alpha }x^{\nu }-T^{\mu \nu
}x^{\alpha }$ is the angular momentum density

with

\begin{equation*}
T^{\alpha \beta }=T_{q}^{\alpha \beta }+T_{g}^{\alpha \beta }=\frac{1}{4}%
\bar{\psi}\gamma \,^{(\alpha }\,i~\overleftrightarrow{D}\,^{\beta )}\psi +%
\frac{1}{4}\left( g^{\alpha \beta }F^{2}-F^{\alpha \mu }F^{\beta }\mu
\right) ~
\end{equation*}

The energy momentum tensor of QCD\ written as the sum of its quark and gluon
parts ($(\alpha \beta )$ means symmetrization in the indices)

\begin{equation}
W_{g}^{\mu \nu \alpha }=W(p^{2})(2ip^{\mu }\gamma ^{\nu }\nu ^{\alpha })+%
\mathrm{other}\text{ }\mathrm{tensor\;}\text{\textrm{structures}}  \label{4}
\end{equation}
$W(p^{2})$ contain nucleon double and single pole contribution as well as a
non singular contribution of the continuum

\begin{equation}
W(p^{2})=\frac{J_{g}\lambda _{N}^{2}}{(p^{2}-m_{N}^{2})^{2}}+\frac{c^{\prime
}}{(p^{2}-m_{N}^{2})}+\cdots  \label{5}
\end{equation}
$\lambda _{N}$ is the coupling of the nucleon to the current $\eta $

\begin{equation}
\langle 0\mid \eta \mid N(p)\rangle =\lambda _{N}~u(p)  \label{6}
\end{equation}

Using a Ward identity the 3-point function(\ref{3}) is rewritten as

\begin{eqnarray}
W_{g}^{\mu \nu \alpha }\underset{\lim q\rightarrow 0}{=}\int dx~dz~z^{\mu
}z^{\nu }\exp (ipx)\exp (iqy)~\langle 0 &\mid &T~\eta (x)~\bar{\eta}%
(0)~O^{\alpha }(z)\mid 0\rangle -\nu \leftrightarrow \alpha  \notag \\
O^{\alpha }(z) &=&\bar{\psi}_{f}~F^{\alpha \beta }~\gamma _{\beta }~\psi (z)
\end{eqnarray}

In the deep Euclidean region $W$ can be calculated in QCD

\begin{eqnarray}
W^{\mathrm{QCD}} &=&\frac{\alpha _{s}}{\pi ^{s}}\left[ \frac{1}{144}\ln
^{2}\left( -\frac{p^{2}}{\mu ^{2}}\right) -\frac{1}{36}\ln \left( -\frac{%
p^{2}}{\mu ^{2}}\right) \right] ~p^{2}  \notag \\
&&-\frac{1}{144\pi ^{2}p^{2}}\langle \frac{\alpha _{s}F^{2}}{\pi }\rangle %
\left[ \ln \left( -\frac{p^{2}}{\mu ^{2}}\right) +\frac{7}{6}-\ln \left( -%
\frac{q^{2}}{\mu ^{2}}\right) \right]  \notag \\
&&-\frac{1}{81\pi p^{4}}\alpha _{s}\langle \bar{q}q\bar{q}q\rangle \left[
20\ln \left( -\frac{p^{2}}{\mu ^{2}}\right) -62\ln \left( -\frac{q^{2}}{\mu
^{2}}\right) \right] +\cdots  \label{8}
\end{eqnarray}

The terms in $\ln \left( -\frac{q^{2}}{\mu ^{2}}\right) $signal the
breakdown of the validity of the OPE for $q^{2}\longrightarrow 0$, they have
to be replaced by bilocal contributions \cite{5}, i.e. one first expands $%
T~\eta (x)\eta (0)=\sum c_{n}(x)O_{n}$ for small values of x, the operators $%
O_{n}$ then combine with $O^{\alpha }(z)$ in a series of two point functions
the leading terms in the asymptotic behaviour of which reproduce the terms
in $\ln \left( -\frac{q^{2}}{\mu ^{2}}\right) $ and which must be
extrapolated to $q^{2}=0$. Expression(\ref{8}) must then be replaced by

\begin{eqnarray}
W^{\,\mathrm{QCD}} &=&\frac{\alpha _{s}}{\pi ^{5}}\left[ \frac{1}{144}\ln
^{2}\left( -\frac{p^{2}}{\mu ^{2}}\right) -\frac{1}{36}\ln \left( -\frac{%
p^{2}}{\mu ^{2}}\right) \right] ~p^{2}  \notag \\
&&-\frac{1}{144\pi ^{2}p^{2}}\langle \frac{\alpha _{s}F^{2}}{\pi }\rangle %
\left[ \ln \left( -\frac{p^{2}}{\mu ^{2}}\right) +\frac{7}{6}\right] -\frac{1%
}{12\pi ^{2}p^{2}}\Pi _{0}(q^{2}=0)  \notag \\
&&+\frac{20}{81\pi p^{4}}\alpha _{s}\langle \bar{q}q^{2}\rangle \ln \left( -%
\frac{p^{2}}{\mu ^{2}}\right) +\frac{4}{3p^{4}}\Pi _{1}(q^{2}=0)  \label{9}
\end{eqnarray}
where 
\begin{equation}
\Pi _{0,1}(q^{2})=\int dz~\exp (iqz)~\langle 0\mid T~O_{5,7}(0)~O^{\alpha
}(z)\mid 0\rangle  \label{10}
\end{equation}
and the relevant operators

\begin{eqnarray}
O_{5}^{\lambda \rho \rho ^{\prime }} &=&2\bar{u}gF^{\lambda \lbrack \rho
}\gamma ^{\rho ^{\prime }]}u-2i\partial ^{\lbrack \rho }(\bar{u}%
\overleftrightarrow{D}^{\lambda }\gamma ^{\rho ^{\prime }]}u)  \notag \\
&&+\bar{u}\overleftarrow{\gamma ^{\mu }D_{\mu }}\,\overrightarrow{D}%
^{\lambda }\,\sigma ^{\rho \rho ^{\prime }}\overleftarrow{D}^{\lambda }%
\overrightarrow{\gamma ^{\mu }D_{\mu }}\,u  \notag \\
&&+\frac{3}{4}\bar{u}gF^{\rho \rho ^{\prime }}\gamma ^{\lambda }u+\frac{3}{4}%
\bar{d}gF^{\rho \rho ^{\prime }}\gamma ^{\lambda }d  \notag \\
O_{7}^{\lambda \rho \rho ^{\prime }} &=&\epsilon ^{ijk}\epsilon ^{i^{\prime
}j^{\prime }k^{\prime }}(D^{\lambda }u)^{i}C\gamma ^{\rho }u^{j}\bar{u}%
^{j^{\prime }}\gamma ^{\rho ^{\prime }}C\bar{u}^{i^{\prime }T}+h.c.
\label{11}
\end{eqnarray}
In the deep Euclidean region

\begin{equation}
\Pi _{0}^{\mathrm{QCD}}(s=q^{2})=-\frac{\alpha _{s}}{60\pi ^{3}}s\ln \left(
-s\right) -\frac{1}{12}\langle \frac{\alpha _{s}F^{2}}{\pi }\rangle \ln
\left( -s\right) +\frac{8}{9}\pi \alpha _{s}\langle \left( \bar{q}q\right)
^{2}\rangle \frac{1}{s}+\cdots  \label{12}
\end{equation}

The method followed here to extract $\Pi _{0}(q^{2}=0)$ differs from the one
used in ref\cite{5}. $\Pi _{0}(s)$ is an analytic function in the complex $s$%
-plane except for cut along the positive real axis. Consider the integral
over the closed contour in the complex $s$-plane consisting of a large
circle of radius $R_{1}$ and two straight lines immediately above and below
the cut which run from threshold to $R_{1}$. By virtue of Cauchy's theorem

\begin{eqnarray}
\frac{1}{2\pi i}\int_{c}\frac{ds}{s}(m^{\prime 2}-s)\Pi _{0}(s) &=&\frac{1}{%
\pi }\int_{th}^{R}\frac{ds}{s}(m^{\prime 2}-s)\func{Im}\Pi _{0}(s)+\frac{1}{%
2\pi i}\oint \frac{ds}{s}(m^{\prime 2}-s)\Pi _{0}(s)  \notag  \label{13} \\
&=&m^{\prime 2}\Pi _{0}(0)
\end{eqnarray}

Where $m^{\prime }$ is a mass parameter. The integral over the real axis
corresponds to hadronic intermediate states with quantum numbers $%
J^{pc}=1^{-~+}$, as $m^{\prime 2}$ is varied from threshold to $R$ this
integral changes sign which implies that it vanishes for some intermediate
value of $m^{\prime 2}$ which is expected to be close to the value of $s$
for which $\func{Im}\Pi _{0}(s)$ reaches its maximum. Experimentally the 
\textbf{E857 }collaboration \cite{8} has recently observed a broad bump in
the invariant mass squared of the $\eta \pi $ system produced in the
reaction $\pi ^{-}p-\longrightarrow \eta \pi ^{-}p$ around $2.0$\textrm{GeV}$%
^{2}$. So if we take $m^{\prime 2}\simeq 2.0$\textrm{GeV}$^{2}$ we expect
the integral over the real axis in eq.(\ref{13}) to be negligible. In the
integral over the circle $\Pi _{0}$ is well approximated by $\Pi _{0}^{%
\mathrm{QCD}}$ except possibly for a small region near the real axis. Eqs.(%
\ref{12}) and (\ref{13}) then yield

\begin{equation}
\Pi _{0}(p^{2}=0)\simeq \frac{\alpha _{s}R_{1}^{2}}{60\pi ^{3}}\left( \frac{%
R_{1}}{3m^{\prime 2}}-\frac{1}{2}\right) +\frac{1}{12}\langle \frac{\alpha
_{s}F^{2}}{\pi }\rangle \left( \frac{R_{1}}{m^{\prime 2}}-\ln \frac{R_{1}}{%
\mu ^{2}}\right) -\frac{8}{9}\pi \alpha _{s}\langle \left( \bar{q}q\right)
^{2}\rangle \frac{1}{m^{\prime 2}}  \label{14}
\end{equation}

In a similar fashion the asymptotic behaviour of $\Pi _{1}(s)$

\begin{equation}
\Pi _{1}^{\mathrm{QCD}}(s)=-\frac{31}{54}\frac{\alpha _{s}}{\pi }\langle 
\bar{q}q^{2}\rangle \ln \left( -\frac{s}{\mu ^{2}}\right) +\langle g\bar{q}%
F\sigma qqq\rangle \left( \frac{1}{3s}\right) +\cdots  \label{15}
\end{equation}
yields 
\begin{equation}
\Pi _{1}(q^{2}=0)=\frac{31}{54}\frac{\alpha _{s}}{\pi }\langle \bar{q}%
q^{2}\rangle \left( \frac{R_{1}}{m^{\prime 2}}-\ln \frac{R_{1}}{\mu ^{2}}%
\right) +\frac{1}{3}\frac{m_{0}^{2}}{m^{\prime 2}}\langle \left( \bar{q}%
q\right) ^{2}\rangle  \label{16}
\end{equation}
where we have parameterized

\begin{equation}
\langle g\bar{q}F\sigma qqq\rangle =-m_{0}^{2}\langle \left( \bar{q}q\right)
^{2}\rangle  \label{17}
\end{equation}
In the equations above $R_{1}\simeq 4$\textrm{GeV}$^{2}$ and $%
m_{0}^{2}\simeq .65$\textrm{GeV}$^{2}$ are reasonable estimates \cite{5}

We are now able to extract the residue $J_{g}$. This is done by considering
the Laplace type integral \cite{9} $\frac{1}{2\pi i}\int_{c}dt~\exp \left( -%
\frac{t}{M^{2}}\right) W(t=p^{2})$ in the complex t-plane over a closed
contour consisting of a circle of radius R and two straight lines above and
below the cut which run from threshold to $R$. $M^{2}$ is the usual 'Borel
mass' parameter. The exponential provides convenient damping of the integral
over the continuum which we expect to be small for an appropriate choice of $%
M^{2}$. If this is done we get a relation between the residues of $W$ and
the integral of $W^{\mathrm{QCD}}$ on the circle of radius $R$%
\begin{eqnarray}
J_{g}+c^{\prime }M^{2} &=&\frac{1}{\lambda _{N}^{2}\exp \left( -\frac{%
m_{N}^{2}}{M^{2}}\right) }\bigg[\frac{\alpha _{s}M^{6}}{36\pi ^{5}}\left( 1-%
\frac{1}{2}\ln \frac{R}{\mu ^{2}}\right) \int_{0}^{\frac{R}{^{M^{2}}}%
}dx~x\exp (-x)  \notag \\
&&+\left[ \frac{1}{144\pi ^{2}}\langle \frac{\alpha _{s}F^{2}}{\pi }\rangle 
\frac{7}{6}+\frac{1}{12\pi ^{2}}\Pi _{0}(0)\right] M^{2}  \notag \\
&&+\frac{1}{144\pi ^{2}}\langle \frac{\alpha _{s}F^{2}}{\pi }\rangle \frac{%
M^{2}}{2\pi i}\oint dt~\exp \left( -\frac{t}{M^{2}}\right) \frac{\ln (-t)}{t}
\notag \\
&&-\frac{20M^{2}}{81\pi }\alpha _{s}\langle \bar{q}q^{2}\rangle \frac{1}{%
2\pi i}\oint \frac{dt}{t^{2}}\exp \left( -\frac{t}{M^{2}}\right) \ln \left(
-t\right) +\frac{4}{3}\Pi _{1}(0)\bigg]  \label{18}
\end{eqnarray}
$\lambda _{N}^{2}$ is obtained from a study of the two point function \cite
{6}

\begin{equation*}
\int dx\,\exp (iqx)\,\langle 0\mid T\,\eta (x)\,\bar{\eta}(0)\mid 0\rangle
\end{equation*}

\begin{eqnarray}
\left( 2\pi \right) ^{4}\lambda _{N}^{2}\,\exp \left( -\frac{m_{N}^{2}}{M^{2}%
}\right) &=&\frac{1}{4}M^{6}\int_{0}^{\frac{R}{M^{2}}}dx\,x^{2}\exp \left(
-x\right) -\frac{\pi ^{2}}{2}\langle \frac{\alpha _{s}F^{2}}{\pi }\rangle
M^{2}\int_{0}^{\frac{R}{M^{2}}}dx\,\exp \left( -x\right)  \notag \\
&&+\frac{32}{3}\pi ^{4}\langle \left( \bar{q}q\right) ^{2}\rangle  \label{19}
\end{eqnarray}

The choice of $M^{2}$ in the equation above and the coherence of the method
is dictated by stability considerations, if there are values of $M^{2}$
small enough to provide adequate damping of the continuum and large enough
to justify the neglect of the contributions of higher order condensates in
the OPE, this should show up in the linear behaviour of the r.h.s. of eq.(%
\ref{18}) in some intermediate range of $M^{2}$. The standard value \cite{9} 
$\langle \frac{\alpha _{s}F^{2}}{\pi }\rangle =0.012$\textrm{GeV}$^{2}$ is
used, for the value of the condensate $\langle (\bar{q}q)^{2}\rangle $ the
choice $\langle \bar{q}q\rangle ^{2}$ (vacuum saturation hypothesis) is
usually made but as it seems too stringent an assumption \cite{10} we take $%
\langle \left( \bar{q}q\right) ^{2}\rangle =\beta \langle \bar{q}q\rangle
^{2}$ the dependance on $\beta $ is weak because it appears both in the
numerator and in the denominator of the r.h.s. of eq.(\ref{18}). In fig.1
the r.h.s. of eq.(\ref{18}) is plotted as a function of $M^{2}$ for $\beta
=1 $ . It is seen that the dependance on $M^{2}$ is essentially linear in
the interval $.7$\textrm{GeV}$^{2}\lesssim M^{2}\lesssim 1.2$\textrm{GeV}$%
^{2}$ with a small value of the slope $(c^{\prime }\simeq 0).$

The result is

\begin{equation}
J_{g}\simeq 0.35\;\;\;\text{for \ \ }\beta =1\;,\;J_{g}\simeq 0.30\;\;\ 
\text{for \ \ }\beta =3  \label{20}
\end{equation}

Balitsky and Ji \cite{5} had reached the same conclusion, they had however
chosen to eliminate the contribution of the nucleon single pole by
multiplying $W$ by $(t-m_{N}^{2})$ before performing the Borel
transformation and used vector meson dominance to estimate $\Pi _{0,1}(0)$.
This has the unfortunate effect of destabilizing the calculation as it can
be seen that the variation of their value for $J_{g}$ with $M^{2}$ is very
rapid.

This can be seen explicitely from the expression obtained following the
method of ref.\cite{5}. Instead of eq.(\ref{18}) one gets 
\begin{eqnarray}
J_{g} &=&\frac{1}{\lambda _{N}^{2}\exp \left( -\frac{m_{N}^{2}}{M^{2}}%
\right) }\bigg[\frac{-\alpha _{s}M^{6}}{36\pi ^{5}}\left( 1-\frac{1}{2}\ln 
\frac{R}{\mu ^{2}}\right) \int_{0}^{\frac{R}{M^{2}}}dx\,x\left( x-\frac{%
m_{N}^{2}}{M^{2}}\right) \exp (-x)  \notag \\
&&+\left[ \frac{1}{144\pi ^{2}}\langle \frac{\alpha _{s}F^{2}}{\pi }\rangle
\cdot \frac{7}{6}+\frac{1}{12\pi ^{2}}\Pi _{0}(0)\right] m_{N}^{2}  \notag \\
&&-\frac{1}{144\pi ^{2}}\langle \frac{\alpha _{s}F^{2}}{\pi }\rangle \cdot 
\frac{1}{2\pi i}\oint dt\exp \left( -\frac{t}{M^{2}}\right) \frac{\ln \left(
-t\right) }{t}(t-m_{N}^{2})  \label{21} \\
&&+\frac{20}{81\pi }\alpha _{s}\langle (\bar{q}q)^{2}\rangle \cdot \frac{1}{%
2\pi i}\oint dt\exp \left( -\frac{t}{M^{2}}\right) \frac{\ln \left(
-t\right) }{t}(t-m_{N}^{2})  \notag \\
&&+\frac{4}{3}\Pi _{1}(0)\left( 1+\frac{m_{N}^{2}}{M^{2}}\right) \bigg] 
\notag
\end{eqnarray}
stability here requires the r.h.s. of eq.(\ref{21}) to be constant in an
intermediate range of $M^{2}$. As can be seen from fig.2 no such thing
happens. This confirms what has been stated above: elimination of the simple
pole contribution by multipication by $(t-m_{N}^{2})$ before perfoming the
Borel multiformation destabilizes the calculation. The main difference
between the approach advocated here and the one of ref.\cite{5}, in addition
to the use of eq.(\ref{13}) instead of vector meson dominance to evaluate
the bilocal contributions $\Pi _{0,1}(0),$consists in not multiplying by $%
(t-m_{N}^{2})$ in order to restore numerical stability.

\bigskip

\bigskip

\textbf{Acknowledgments:} I thank the Academie de Versailles for its
generous support and the IPN, Orsay, where the essential part of this work
was done, for their kind hospitality

\pagebreak

{\Large Figure Captions:}

\bigskip

{\large fig.1: }The r.h.s.of eq.(\ref{18}) plotted against $M^{2}$. The
linear variation of

$J_{g}+c^{\prime }M^{2}$ is apparent in the range $.7\mathrm{GeV}%
^{2}\lesssim M^{2}\lesssim 1.2\mathrm{GeV}^{2}$

\bigskip

{\large fig.2:} The r.h.s. of eq.(\ref{21}) plotted against $M^{2}$. $J_{g}$
which ought to be constant in some range of $M^{2}$ varies rapidly.

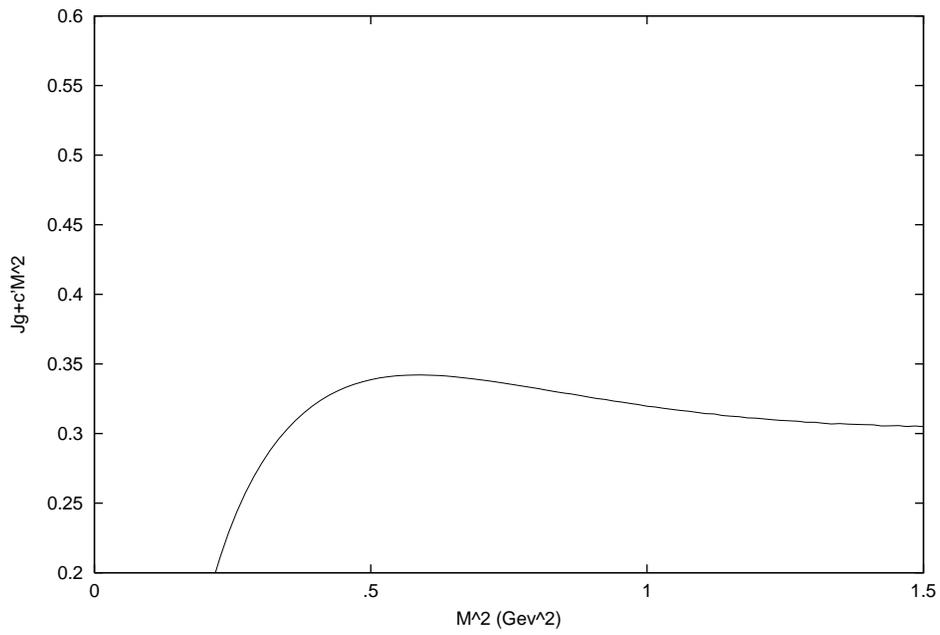
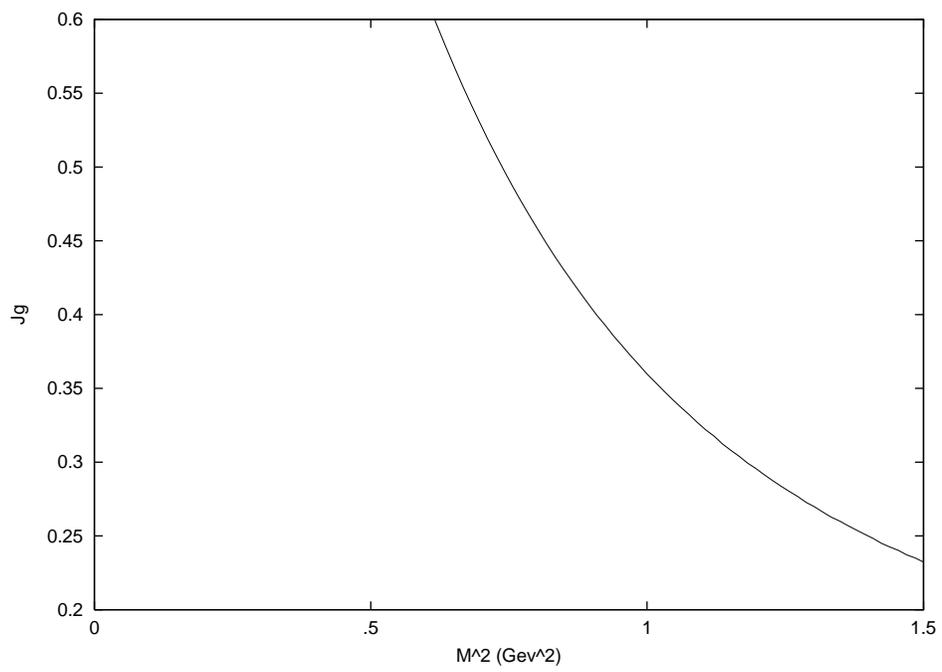
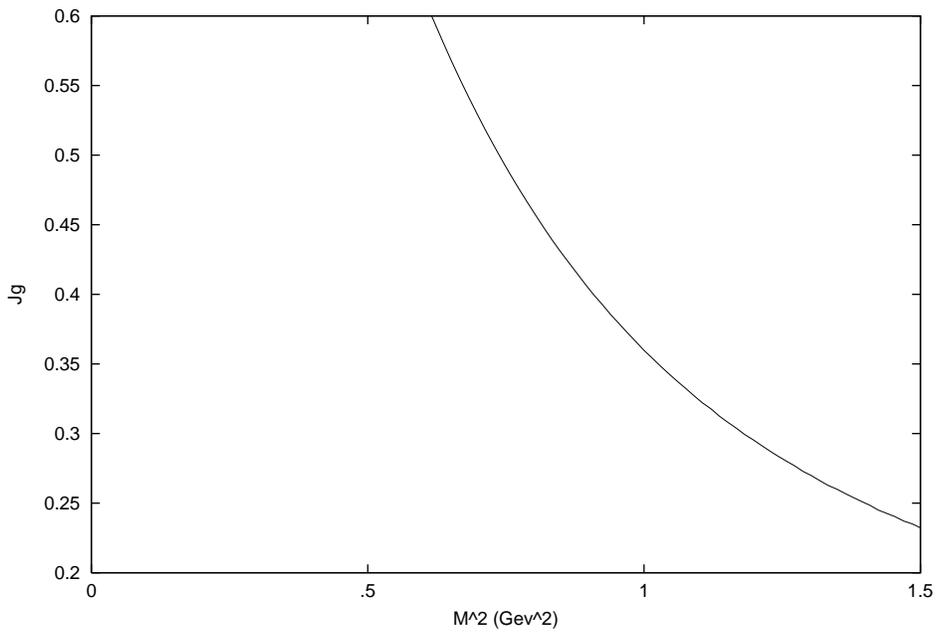
\end{document}